# Under-approximation of the Greatest Fixpoint in Real-Time System Verification[*]

Farn Wang

Dept. of Electrical Engineering, National Taiwan University
1, Sec. 4, Roosevelt Rd., Taipei, Taiwan 106, ROC;
+886-2-23635251 ext. 435; FAX:+886-2-23671909;
farn@cc.ee.ntu.edu.tw; http://cc.ee.ntu.edu.tw/∼farn

Model-checker/simulator **RED** 5.4 and benchmarks are available at
http://cc.ee.ntu.edu.tw/∼val

**Abstract.** Techniques for the efficient successive under-approximation of the greatest fixpoint in TCTL formulas can be useful in fast refutation of inevitability properties and vacuity checking. We first give an integrated algorithmic framework for both under and over-approximate model-checking. We design the *NZF (Non-Zeno Fairness) predicate*, with a greatest fixpoint formulation, as a unified framework for the evaluation of formulas like $\exists\Box\eta_1$, $\exists\Box\Diamond\eta_1$, and $\exists\Diamond\Box\eta_1$. We then prove the correctness of a new formulation for the characterization of the NZF predicate based on zone search and the least fixpoint evaluation. The new formulation then leads to the design of an evaluation algorithm, with the capability of successive under-approximation, for $\exists\Box\eta_1$, $\exists\Box\Diamond\eta_1$, and $\exists\Diamond\Box\eta_1$. We then present techniques to efficiently search for the zones and to speed up the under-approximate evaluation of those three formulas. Our experiments show that the techniques have significantly enhanced the verification performance against several benchmarks over exact model-checking.

**Keywords:** real-time, model-checking, verification, fairness, under-approximation, greatest fixpoint, inevitability, vacuity

## 1 Introduction

The greatest fixpoint evaluation is useful in the verification of *inevitability properties* and *fairness assumptions* of real-time systems [19,21]. In the before, people have researched on *over-approximation* techniques [12,21], that construct the characterization for a superset of the greatest fixpoint. But in practice, *under-approximation* techniques, that construct a subset of the greatest fixpoint, can also be useful. For example, we may want to verify a *TCTL (Timed Computation*

---

[*] The work is partially supported by NSC, Taiwan, ROC under grants NSC 92-2213-E-002-103, NSC 92-2213-E-002-104, and by the System Verification Technology Project of Industrial Technology Research Institute, Taiwan, ROC (2004).

*Tree Logic)* [2] inevitability property like $\forall\Diamond\mathtt{full}$ which says *"Along all computations, our stomachs will eventually be full of nice food."* In model-checking, we actually try to prove the falsehood of its negation, or equivalently the emptiness of the state space characterized by $\exists\Box\neg\mathtt{full}$. The evaluation procedure for the greatest fixpoints in TCTL in exact analysis is quite expensive [21]. In real-world system development, it is seldom the case that no design bug ever happens. On the contrary, it is quite possible that a long debugging process is needed in the early design stages. Thus not only we need techniques to prove correct inevitabilities, but also we are in need of techniques to fast refute incorrect inevitabilities. Efficient under-approximation techniques can serve to this purpose by quickly constructing a few *counter-examples* to the incorrect inevitabilities to make fast refutation.

Under-approximation techniques for the greatest fixpoint evaluation can also be useful in the *vacuity checking* [4,9] of system assumptions. For example, we may want to specify that *"Whenever we are hungry, along all computations, eventually we will be full."* In TCTL, this is $\forall\Box(\mathtt{hungry} \to \forall\Diamond\mathtt{full})$. But this property can be satisfied by a system model that either does not generate any computation or never gets into a $\mathtt{hungry}$ state. In the end, it is still the responsibility of the verification engineers to check if a specification is vacuously satisfied because of the wrong modeling of the environment or the component interactions. Under-approximation can help in this case by quickly constructing a few example computations, if any, along which eventually $\mathtt{hungry}$ is true.

Similarly, vacuous satisfaction can happen in the verification of properties with fairness assumptions. For example, we may specify that *"If we are full infinitely many times, then we eventually will not feel hungry."* Again, the property can be vacuously satisfied in a system that fills us only finitely many times. Before evaluating the specification, we may first want to use under-approximation techniques to make sure that in the model, there are computations with infinitely many $\mathtt{full}$ states.

In this work, we present techniques for the under-approximation of the greatest fixpoint evaluation in real-time systems with fairness assumptions. We propose a predicate called *NZF (Non-Zeno Fairness)* as a unified framework for the evaluation of the greatest fixpoints for formulas like $\exists\Box\phi$, $\exists\Box\Diamond\phi$[1], and $\exists\Diamond\Box\phi$[2] in a TCTL extension with the fairness concepts. In notations, the predicate is $NZF(\eta_0, \eta_1, \eta_2)$. A state $\nu_0$ satisfies $NZF(\eta_0, \eta_1, \eta_2)$ iff $\nu_0$ starts a run along which $\eta_0$ is always true, $\eta_1$ is eventually always true, and $\eta_2$ is true infinitely often. The evaluation of $NZF(\eta_0, \eta_1, \eta_2)$ consists of all states that go (through a path first satisfying $\eta_0$ and then satisfying $\eta_1$) to some fair computation cycles such that the execution time along each cycle is no less than 1, $\eta_0 \wedge \eta_1$ is always true along the cycles, and $\eta_2$ is true at least once along each cycle. For convenience, we call such a cycle an $(\eta_1, \eta_2)$-*NZF-cycle*. A picture showing the states, run seg-ments, and cycles in the evaluation of an $NZF(\eta_0, \eta_1, \eta_2)$ predicate is in figure 1. According to [21], the evaluation of $NZF()$ needs a greatest fixpoint evaluation loop nesting a least fixpoint evaluation loop and is quite expensive.

---

[1] This means that there is an infinite computation along which $\phi$ is true infinitely many times.
[2] This means that there is an infinite computation along which $\phi$ eventually becomes true forever.

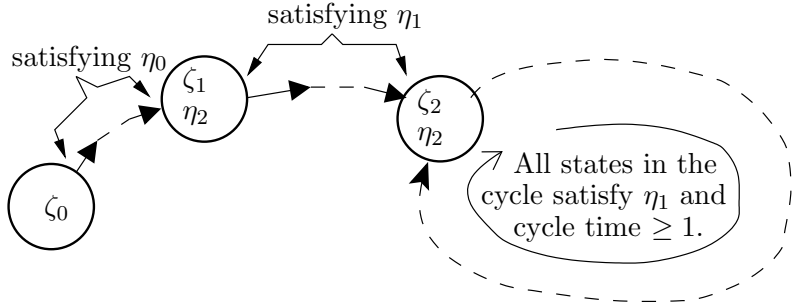

**Fig. 1.** The run segments for $NZF(\eta_0, \eta_1, \eta_2)$

Given two set $\eta_1, \eta_2$ of states, let $\texttt{rch\_bck}(\eta_1, \eta_2)$ be the predicate characterizing those states backwardly reachable from $\eta_2$ through paths of states in $\eta_1$. In the literature, $\texttt{rch\_bck}()$ is evaluated as a least-fixpoint. (Details in section 5.) In this work, we propose the following new formulation for its evaluation:

$$NZF(\eta_0, \eta_1, \eta_2) \equiv \bigvee_{\zeta \text{ is a set of states in an } (\eta_1, \eta_2)\text{-NZF-cycle.}} \texttt{rch\_bck}(\eta_0, \texttt{rch\_bck}(\eta_1, \zeta))$$

The formulation is based on the enumeration of state sets in NZF-cycles and only two least fixpoint evaluations for each state set in the enumeration. There are two advantages to this formulation.

- It allows for successive under-approximation. After any iteration, if the verification engineers see that either enough precision has been reached or too much computation resources have been consumed, she/he can terminate the enumeration.
- It may seem that the policy of the NZF zone enumeration can greatly affect the efficiency to approach the fixpoint solution. In reality, our experiment data shows that the performance of this formulation can be very insensitive to the enumeration policy. As long as the NZF states fall in the backward reachability of a zone, it will be included in under-approximation. In a well-designed system, usually it is the case that most of the states are reachable from one another. In particular, we have established lemma 6 to show that in each iteration of the under-approximation, states in an NZ-cycle either all will be included in the under-approximation or none will.

Since our new formulation can be insensitive to the policy of NZF-cycle state set enumeration, it is better to first enumerate those NZF-cycle state sets that can be efficiently constructed. We have developed two techniques for quick construction of the NZF-cycle state sets. Our experiment shows that our implementation can lead up to $1000^+$ times speed-up against some of the benchmarks. Moreover, in most cases, we succeeded in refuting the inevitabilities or proving vacuities after enumerating only one or two state sets.

Section 2 discusses related work. Section 3 reviews the mathematical models of our system behaviors. Section 4 extends TCTL [2] to $TCTL^\infty$ to allow for the specification of fairness properties. Section 5 gives the background knowledge for reading this article. Then section 6 presents an algorithmic framework for both over-approximate and under-approximate model-checking of $TCTL^\infty$ properties. Section 7 explains why the old formulation for NZF evaluation [21] is expensive. Section 8 presents our successive under-approximation algorithm, including the new theoretical formulation of the NZF evaluation and two techniques to construct characterizations of those states in the NZF-cycles. Section 9 reports a speed-up technique, for our under-approximation algorithm, that does not sacrifice the precision of the greatest fixpoint evaluation. Section 10 reports our implementation and experiments. The experiment data shows significant enhancement over the exact analysis against several benchmarks.

## 2  Related work

The model of *timed automata (TA)* was by Alur and Dill [3]. TCTL and its model-checking algorithm was by Alur, Cocoubetis, and Dill [2]. Symbolic model-checking algorithm based on zones was by Henzinger et al [8].

Wong-Toi presented a general framework for the approximate verification of TAs [22]. Especially, the convex-hull over-approximation has been shown a very powerful technique in many following workpieces.

Moller applied over-approximation techniques to analyze restricted TCTL inevitability properties without modal-formula nesting [12]. The idea was to make model augmentations to speed up the verification performance.

Wang and et al discussed how to speed up the greatest fixpoint evaluation in TCTL [21]. Other than the over-approximation, they also presented a speed-up technique called *EDGF (Early Decision on Greatest Fixpoint)* which yields exact analysis results. The idea is that inevitability analysis usually takes the form of $\forall\Box(p \rightarrow \forall\Diamond q)$, which after negation for model-checking, becomes the reachability of $p \wedge \exists\Box\neg q$. While evaluating the greatest fixpoint of $\exists\Box\neg q$, the image of the greatest fixpoint monotonically shrinks in successive iterations. Thus when we find the intersection between $p$ and the image is empty at a particular iteration, we can terminate the fixpoint evaluation rightaway. Note that EDGF speeds up the verification only when an inevitability is correct. When an inevitability is incorrect, it shows no performance enhancement.

Wang extended TCTL with the capability of punctual event specifications and multiple strong and weak fairness assumptions [19]. The evaluation algorithm of those fairness assumptions was based on the greatest fixpoint evaluation.

## 3 Timed automatas

We use the widely accepted model of *timed automata (TA)* [3] to describe the transitions in dense-time state-spaces. A TA is a finite-state automata equipped with a finite set of clocks which can hold nonnegative real-values. At any moment, the TA can stay in only one *mode* (or *control location*). Each mode is labeled with an invariance condition on clocks. At any instant, at most one transition can be fired if its triggering condition is satisfied. Upon the firing, the automata instantaneously transits from one mode to another and resets some clocks to zero. In between transitions, all clocks increase their readings at a uniform rate.

For convenience, given a set $P$ of atomic propositions and a set $X$ of clocks, we use $B(P, X)$ as the set of all Boolean combinations of atoms of the forms $p$ and $x \sim c$ where $p \in P$, $x \in X \cup \{0\}$, "$\sim$" is one of $\leq, <, =, >, \geq$, and $c$ is an integer constant.

**Definition 1. <u>timed automata (TA)</u>** A TA $A$ is given as a tuple $\langle X, Q, I, \mu, E, \gamma, \tau, \pi \rangle$ with the following restrictions. $X$ is a finite set of clocks. $Q$ is a finite set of modes. $I \in B(Q, X)$ is the initial condition. $\mu : Q \mapsto B(\emptyset, X)$ defines the conjunctive invariance condition of each mode. $E$ is the finite set of transitions. $\gamma : E \mapsto (Q \times Q)$ defines the source and the destination modes of each transition. $\tau : E \mapsto B(\emptyset, X)$ and $\pi : E \mapsto 2^X$ respectively defines the conjunctive triggering condition and the clock set to reset of each transition. ■

**Definition 2. <u>states</u>** A state $\nu$ of TA $A = \langle X, Q, I, \mu, E, \gamma, \tau, \pi \rangle$ is a valuation from $Q \cup X$ such that for all $q \in Q$, $\nu(q) \in \{\textit{false}, \textit{true}\}$ and for all $x \in X$, $\nu(x) \in \mathcal{R}^+$, the set of nonnegative real numbers. The restriction is that there is at most one $q \in Q$ such that $\nu(q)$ is *true*. ■

**Definition 3. <u>Satisfaction of state predicates</u>** We say a state $\nu$ satisfies a state predicate $\eta \in B(P, X)$, where $P$ is either $\emptyset$ or $Q$, iff the following inductive conditions are satisfied.
- $\nu \models q$ iff $\nu(q)$;
- $\nu \models x \sim c$ iff $\nu(x) \sim c$;
- $\nu \models \eta_1 \vee \eta_2$ iff $(q, \nu) \models \eta_1$ or $(q, \nu) \models \eta_2$;
- $\nu \models \neg \eta_1$ iff it is not the case that $\nu \models \eta_1$. ■

For any $\delta \in \mathcal{R}^+$, $\nu + \delta$ is a new state identical to $\nu$ except that for every $x \in X$, $\nu(x) + \delta = (\nu + \delta)(x)$. Given $\bar{X} \subseteq X$, $\nu\bar{X}$ is a new state identical to $\nu$ except that for every $x \in \bar{X}$, $\nu\bar{X}(x) = 0$. Given $q \in Q$, $\nu q$ is identical to $\nu$ except that $\nu(q) = \textit{true}$.

**Definition 4. <u>runs</u>** Given a TA $A = \langle X, Q, I, \mu, E, \gamma, \tau, \pi \rangle$, a *run* is an infinite computation of $A$ along which time diverges. Formally speaking, a run is an infinite sequence of state-time pairs $(\nu_0, t_0)(\nu_1, t_1) \ldots (\nu_k, t_k) \ldots \ldots$ such that
- $t_0 t_1 \ldots t_k \ldots \ldots$ is a monotonically increasing divergent real-number sequence, i.e., $\forall c \in \mathcal{N}, \exists h > 1, t_h > c$; and
- **Invariance condition:** for all $k \geq 0$, $\delta \in [0, t_{k+1} - t_k]$, and $q \in Q$, $\nu_k + \delta \models \mu(q)$ iff $\nu_k \models \mu(q)$; and

- **Transitions:** for all $k \geq 0$, either
  - **a null transition happens**, i.e., $\nu_k + (t_{k+1} - t_k) = \nu_{k+1}$; or
  - **a discrete transition $e$ happens**, denoted $q_k \xrightarrow{e} q_{k+1}$ for some $e \in E$ such that $\gamma(e) = (q_k, q_{k+1}) \wedge \nu_k \models \mu(q_k) \wedge \nu_{k+1} \models \mu(q_{k+1})$. The constraint is that $\nu_k + t_{k+1} - t_k \models \tau(e)$, and $(\nu_k + t_{k+1} - t_k)\pi(e)q_{k+1} = \nu_{k+1}$. ∎

## 4  TCTL$^\infty$

$TCTL^\infty$ is an extension of TCTL [2] and has the following syntax rules.

$$\phi ::= q \,|\, x \sim c \,|\, \phi_1 \vee \phi_2 \,|\, \neg \phi_1 \,|\, x.\phi_1 \,|\, \exists \phi_1 \mathcal{U} \phi_2 \,|\, \exists \Box \phi_1 \,|\, \exists \Box \Diamond \phi_1 \,|\, \exists \Diamond \Box \phi_1$$

Here $q \in Q$, $x \in X$, $c \in \mathcal{N}$. $\phi_1$ and $\phi_2$ are $TCTL^\infty$ formulas. $x.\phi$ means that if there is a clock $x$ with reading zero now, then $\phi$ is satisfied. $\exists \phi_1 \mathcal{U} \phi_2$ means that there exists a computation along which $\phi_1$ is true until $\phi_2$ happens. $\exists \Box \phi_1$ means that there is a computation along which $\phi_1$ is true in every state. $\exists \Box \Diamond \phi_1$ means that there is a computation along which $\phi_1$ is true infinitely often. $\exists \Diamond \Box \phi_1$ means that there is a computation along which $\phi_1$ will be stable eventually. Sometimes, $\Box \Diamond$ is written as $\Diamond^\infty$ while $\Diamond \Box$ as $\Box^\infty$ in the literature. Also we adopt the following standard shorthands :

- *true* for $0 = 0$
- $\phi_1 \wedge \phi_2$ for $\neg((\neg \phi_1) \vee (\neg \phi_2))$
- $\exists \Diamond \phi_1$ for $\exists \mathit{true} \mathcal{U} \phi_1$
- $\forall \phi_1 \mathcal{U} \phi_2$ for $\neg((\exists (\neg \phi_2) \mathcal{U} \neg (\phi_1 \vee \phi_2)) \vee (\exists \Box \neg \phi_2))$
- $\forall \Box \Diamond \phi_1$ for $\neg \exists \Diamond \Box \neg \phi_1$
- *false* for $\neg \mathit{true}$
- $\phi_1 \rightarrow \phi_2$ for $(\neg \phi_1) \vee \phi_2$
- $\forall \Box \phi_1$ for $\neg \exists \Diamond \neg \phi_1$
- $\forall \Diamond \phi_1$ for $\forall \mathit{true} \mathcal{U} \phi_1$
- $\forall \Diamond \Box \phi_1$ for $\neg \exists \Box \Diamond \neg \phi_1$

**Definition 5. (Satisfaction of $TCTL^\infty$ formulas):** We write in notations $A, \nu_1 \models \phi$ to mean that $TCTL^\infty$ formula $\phi$ is satisfied at state $\nu_1$ in TA $A$. The satisfaction relation is defined inductively as follows.
- When $\eta \in B(Q, X)$, $A, \nu_1 \models \eta$ iff $\nu_1 \models \eta$, which was previously defined in the beginning of section 3.
- $A, \nu_1 \models \phi_1 \vee \phi_2$ iff either $A, \nu_1 \models \phi_1$ or $A, \nu_1 \models \phi_2$.
- $A, \nu_1 \models \neg \phi_1$ iff $A, \nu_1 \not\models \phi_1$.
- $A, \nu_1 \models x.\phi$ iff $A, \nu_1\{x\} \models \phi$.
- $A, \nu_1 \models \exists \phi_1 \mathcal{U} \phi_2$ iff there exists a run $\rho = (\nu_1, t_1)(\nu_2, t_2)\ldots$ and an $i \geq 1$ such that $A, \nu_i \models \phi_2$ and for every $0 \leq j < i$ and $\delta \in [0, t_{j+1} - t_j]$, $A, \nu_j + \delta \models \phi_1$.
- $A, \nu_1 \models \exists \Box \phi_1$ iff there exists a run $\rho = (\nu_1, t_1)(\nu_2, t_2)\ldots$ such that for every $i \geq 1$ and $\delta \in [0, t_{i+1} - t_i]$, $A, \nu_i + \delta \models \phi_1$.
- $A, \nu_1 \models \exists \Box \Diamond \phi_1$ iff there exist a run $\rho = (\nu_1, t_1)(\nu_2, t_2)\ldots$ and an infinite and divergent positive integer sequence $i_1 i_2 \ldots i_k \ldots$ such that for every $k \geq 1$, $A, \nu_{i_k} \models \phi_1$.
- $A, \nu_1 \models \exists \Diamond \Box \phi_1$ iff there exist a run $\rho = (\nu_1, t_1)(\nu_2, t_2)\ldots$ and $k \geq 1$ such that for every $i \geq k$ and $\delta \in [0, t_{i+1} - t_i]$, $A, \nu_i + \delta \models \phi_1$.

A TA $A = \langle X, Q, I, \mu, E, \gamma, \tau, \pi \rangle$ satisfies a $TCTL^\infty$ formula $\phi$, in symbols $A \models \phi$, iff for every state $\nu_0 \models I$, $A, \nu_0 \models \phi$. ∎

## 5   Basic building blocks from the literature

For convenience, let $\mathcal{Z}$ be the set of integers. Given $c \geq 0$ and $c \in \mathcal{Z}$, let $\mathcal{I}_c$ be $\{\infty\} \cup \{d \mid d \in \mathcal{Z}; -c \leq d \leq c\}$. Also let $C_{A:\phi}$ be the biggest timing constant used in TA $A$ and $TCTL^\infty$ formula $\phi$.

Most modern model-checkers for timed systems are built around some symbolic manipulation procedures [8] of *zones* implemented in various data-structures [1,5,6,10,11,14,15]. A zone is symbolically represented by a set of difference constraints between clock pairs. Formally, a zone is a conjunction of constraints like $x - x' \sim d$, with $x, x' \in X \cup \{0\}$, $\sim \in \{\text{``}\leq\text{''}, \text{``}<\text{''}\}$, and $d \in \mathcal{I}_{C_{A:\phi}}$, such that when $d = \infty$, $\sim$ must be "<". For convenience, let $\mathcal{B}_c = \{(\sim, d) \mid \sim \in \{\text{``}\leq\text{''}, \text{``}<\text{''}\}; d \in \mathcal{I}_c; d = \infty \Rightarrow \sim = \text{``}<\text{''}\}$. With respect to given $X$ and $C_{A:\phi}$, the set of all zones is finite. Alternatively, a zone can be defined as a mapping $(X \cup \{0\})^2 \mapsto \mathcal{B}_{C_{A:\phi}}$. We shall use the two equivalent notations flexibly.

We need two basic procedures, one for the computation of weakest preconditions of discrete transitions and the other for those of backward time-progressions. Details about the two procedures can be found in [8, 14–17, 20]. Given a state-space representation $\eta$ (as a union of zones) and a discrete transition $e$, the first procedure, $\texttt{xtion\_bck}(\eta, e)$ with $\gamma(e) = (q, q')$, computes the weakest precondition

- in which every state satisfies the invariance condition $\mu(q)$; and
- from which we can transit to states in $\eta$ through $e$.

$\eta$ can be represented as a DBM set [6] or as a BDD-like data-structure [14,16,18]. Our algorithms are independent of the representation scheme of $\eta$. The second procedure, $\texttt{time\_bck}(\eta)$, computes the space representation of states

- from which we can go to states in $\eta$ simply by time-passage; and
- every state in the time-passage also satisfies the invariance condition imposed by $\mu()$ for whatever modes the states are in.

We can go from zone $\eta_1$ to another zone $\eta_2$ in one time-progress step iff $\eta_1 \subseteq \texttt{time\_bck}(\eta_2)$ and in one discrete transition step iff $\exists e \in E(\eta_1 \subseteq \texttt{xtion\_bck}(\eta_2, e))$. A zone sequence $\eta_1 \eta_2 \ldots \eta_k$ corresponds to a finite segment of computation iff for every $1 \leq i < k$, either $\eta_i \subseteq \texttt{time\_bck}(\eta_{i+1})$ or $\eta_i \subseteq \texttt{xtion\_bck}(\eta_{i+1})$.

With the two basic procedures, we can construct the symbolic backward reachability procedure, denoted $\texttt{rch\_bck}(\eta_1, \eta_2)$ for convenience, as in [8,14–17, 20]. Intuitively, $\texttt{rch\_bck}(\eta_1, \eta_2)$ characterizes the state-space for $\exists \eta_1 \mathcal{U} \eta_2$. Computationally, $\texttt{rch\_bck}(\eta_1, \eta_2)$ can be defined as the least fixpoint of equation: $F = \eta_2 \vee \big(\eta_1 \wedge \texttt{time\_bck}(\eta_1 \wedge \bigvee_{e \in \mathsf{E}} \texttt{xtion\_bck}(F, e))\big)$. That is, $\texttt{rch\_bck}(\eta_1, \eta_2) \equiv \texttt{lfp} F. \big(\eta_2 \vee \big(\eta_1 \wedge \texttt{time\_bck}(\eta_1 \wedge \bigvee_{e \in \mathsf{E}} \texttt{xtion\_bck}(F, e))\big)\big)$. The least fixpoint is computable because of the monotonicity of $F$ in fixpoint equation $F = \Lambda(F)$ and the finite structure of a zone space.

To calculate the weakest precondition before a clock reset, we also need a partial implementation of the Fourier-Motzkin elimination [7]. We assume that we have such a procedure $\texttt{FM\_elim}(\eta, \{x\})$ which eliminates all information in state-predicate $\eta$ related to $x$.

## 6 Abstract model-checking algorithm

The key component in our abstract model-checking algorithm is for the approximate construction of the symbolic representations of states that satisfy one of the following three types of properties: $\exists\Box\phi_1$, $\exists\Box\Diamond\phi_1$, and $\exists\Diamond\Box\phi_1$. We can now establish the following lemmas within the context of a given state $\nu$ of a TA $A$. Due to page-limit, the proofs are omitted.

**Lemma 1.** $A, \nu \models \exists\Box\psi_1$ *iff* $A, \nu \models \text{NZF}(\psi_1, \psi_1, \text{true})$. ∎

**Lemma 2.** $A, \nu \models \exists\Box\Diamond\psi_1$ *iff* $A, \nu \models \text{NZF}(\text{true}, \text{true}, \psi_1)$. ∎

**Lemma 3.** $A, \nu \models \exists\Diamond\Box\psi_1$ *iff* $A, \nu \models \text{NZF}(\text{true}, \psi_1, \text{true})$. ∎

The three lemmas together show that the evaluation of $NZF(\eta_0, \eta_1, \eta_2)$ can be used as a unified scheme to evaluate these three types of the properties.

We are about to give a framework of abstract model-checking algorithm that is capable of doing both under and over-approximations. In general, over-approximation can be done in various ways [21,22], e.g. the convex-hull approximation. In the following sections, we focus on how to do under-approximation of $NZF()$. For the time being, we assume that there is a procedure called $\text{NZF}(\eta_0, \eta_1, \eta_2, \text{flag}_A)$ capable of calculating the approximation of $NZF(\eta_0, \eta_1, \eta_2)$. Here, input flag $\text{flag}_A$ is used to choose the approximation scheme. When $\text{flag}_A$ is -1, it means under-approximation; 0, no approximation; and 1, over-approximation.

The following evaluation algorithm uses $\text{NZF}()$ and the procedures presented in the last section as basic blocks to evaluate $TCTL^\infty$ formulas.

```
eval(A, φ̄, flag_A) {
   switch (φ̄) {
   case (false):      return false;
   case (q):          return q;
   case (x ∼ c):      return x ∼ c; ;
   case (φ_1 ∨ φ_2):  return val(A, φ_1, flag_A) ∨ eval(A, φ_2, flag_A);
   case (¬φ_1):       return ¬eval(A, φ_1, −1 ∗ flag_A);
   case (x.φ_1):      return FM_elim(x = 0 ∧ eval(A, φ_1, flag_A), {x});
   case (∃φ_1 U φ_2): Y_1 := eval(A, φ_1, flag_A);
                      Y_2 := eval(A, φ_2, flag_A) ∧ NZF(true, true, true, flag_A);
                      return rch_bck(Y_1, Y_2);
   case (∃□φ_1):      W := eval(A, φ_1, flag_A); return NZF(W, W, true, flag_A));
   case (∃□◇φ_1):     W := eval(A, φ_1, flag_A); return NZF(true, true, W, flag_A));
   case (∃◇□φ_1):     W := eval(A, φ_1, flag_A); return NZF(true, W, true, flag_A));
   }
}
```

The correctness of the algorithm can be established with the following lemma.

**Lemma 4.** *Suppose we have a correct implementation for* $\text{NZF}(\eta_0, \eta_1, \eta_2, \text{flag}_A)$. *Procedure* $\text{eval}(A, \bar{\phi}, \text{flag}_A)$ *yields an under-approximation* of $\bar{\phi}$ when $\text{flag}_A = -1$; an exact evaluation when $\text{flag}_A = 0$; and an over-approximation of $\bar{\phi}$ when $\text{flag}_A = 1$. ∎

Then in exact analysis of model-checking, TA $A$ satisfies $TCTL^\infty$ formula $\phi$ iff $I \wedge \text{eval}(A, \neg\phi, 0)$ is *false*. In over-approximation, $A$ does not satisfy $\phi$ if $I \wedge \text{eval}(A, \neg\phi, -1)$ is not *false*. In under-approximation, $A$ satisfies $\phi$ if $I \wedge \text{eval}(A, \neg\phi, 1)$ is *false*.

## 7 Formulation for NZF evaluation in the literature

In the following, we discuss paths and cycles of zones. A finite zone path along which all zones satisfy $\eta_1$ is called an $\eta_1$-*path*. Infinite computations can be formulated as *non-Zeno cycles* (or strongly connected components) of zones such that the execution times of the cycles are at least 1.

According to [19], a formulation of $NZF(\eta_0, \eta_1, \eta_2)$ evaluation is

$$\text{rch\_bck}(\eta_0, \text{rch\_bck}(\eta_1, \text{gfp}F.\exists z.(z = 0 \wedge \eta_2 \wedge \text{rch\_bck}(\eta_1, z \geq C_{A:\phi} \wedge F)))) \quad (1)$$

Here $\text{gfp}$ means the greatest fixpoint. $\text{gfp}F.\Lambda(F)$ specifies the greatest fixpoint of $\Lambda()$. Clock $z$ is used to check if the cycle time is no less than 1. Formulation (1) involves a double loop and two single loops in implementation.[3] The double loop evaluates the $(\eta_1, \eta_2)$-NZF-cycles and is as follows.

$$\text{gfp}F.\exists z.(z = 0 \wedge \eta_2 \wedge \text{rch\_bck}(\eta_1, z \geq 1 \wedge F)) \quad (2)$$

The inner loop (i.e., $\text{rch\_bck}(\eta_1, z \geq 1 \wedge F)$) of the double loop is for the least fixpoint evaluation of an $\eta_1$-path that starts with a zone satisfying $\eta_2$. The outer loop of the double loop is for the greatest fixpoint evaluation of an NZF-cycle through checking whether the starting and the ending zones of the $\eta_1$-path yielded from the inner cycle coincide. Then after the double loop has been executed, the two outmost invocations of $\text{rch\_bck}()$ in (1) incurs two single-loop executions to calculate a predicate that characterizes every zone $\zeta_0$ in figure 1, i.e., the set of states starting an $\eta_0$-path to zone $\zeta_1$ that starts an $\eta_1$-path to an $(\eta_1, \eta_2)$-NZF-cycles.

As reported in [21], the double loop in formulation (1) dominates the complexity and can result in very expensive computations.

## 8 Successive under-approximation of *NZF*()

In the following, we first present a new formulation for the NZF evaluation based on zone searching and least fixpoint evaluation. Then we propose techniques for zone searching in subsections 8.2 and 8.3. Finally, we integrate the ingredients to present a successive under-approximation algorithm for *NZF*().

---

[3] An invocation of $\text{rch\_bck}()$ is executed as a loop. An evaluation of the greatest fixpoint also incurs a loop execution.

### 8.1 Another formulation of NZF evaluation

The basic idea of our under-approximate evaluation of *NZF*() is the following. If somehow we know that a particular zone $\zeta_1$ is in an $(\eta_1, \eta_2)$-NZF-cycle, then we can readily do `rch_bck`$(\eta_0, $`rch_bck`$(\eta_1, \zeta_1))$ to characterize a set of states that satisfy $NZF(\eta_0, \eta_1, \eta_2)$. If we can make a good guess for such $\zeta_1$, then it is possible that `rch_bck`$(\eta_0, $`rch_bck`$(\eta_1, \zeta_1))$ could turn out to be a big chunk of the exact evaluation of $NZF(\eta_0, \eta_1, \eta_2)$. Then a few such good guesses could give us a very precise under-approximation of the greatest fixpoint. The idea can be formalized as the following new formulation of *NZF*() evaluation. For convenience, a zone is called an $(\eta_1, \eta_2)$-*NZF-zone* if it is in an $(\eta_1, \eta_2)$-NZF-cycle.

**Lemma 5.** $\text{NZF}(\eta_0, \eta_1, \eta_2) \equiv \bigvee_{\zeta \text{ is an } (\eta_1, \eta_2)\text{-NZF-zone.}} \texttt{rch\_bck}(\eta_0, \texttt{rch\_bck}(\eta_1, \zeta))$. ∎

Proof of the lemma can be found in appendix A.

### 8.2 Fair zones without upper bounds in the cycle

First, we can check those zones characterized by $\eta_1 \wedge \eta_2$. If there are such zones that have no upper bounds on all clock readings, then these zones constitute self-cycle with arbitrary execution times. Such a zone, say $\zeta$, are characterized by the following condition: $\forall x \in X, \zeta(x, 0) = (<, \infty)$. The argument for guessing the existence of such zones is that in many embedded systems, for example the communication protocols, there usually are the idle modes which the systems repeatedly enter after a communication session. In such idle modes, the systems usually wait without time-bounds until some events happen. Thus it is very likely that such idle modes may exist in the system descriptions. Furthermore, we do not have to run the expensive double loop in formulation (1) to check the non-Zenoness of the self-cycle. As we will see in the experiment report, this strategy indeed can alone lead to under-approximation precise enough for the refutation of many benchmarks.

Assume that $\eta_1 \wedge \eta_2 = \bigvee_{1 \leq k \leq n} \zeta_k$, i.e., a disjunction of $n$ zones, the union of all zones without upper bounds on all clocks in $\eta_1 \wedge \eta_2$ can be constructed with the following procedure.

$$\texttt{get\_zones\_wo\_upperbounds}(\zeta_1 \vee \ldots \vee \zeta_n) \equiv \bigvee_{1 \leq k \leq n; \forall x \in X, \zeta_k(x,0)=(<,\infty)} \zeta_k$$

### 8.3 Searching for non-Zeno cycles from fair zones

`get_zones_wo_upperbounds`() represents an efficient way to find some specific zones in the $(\eta_1, \eta_2)$-NZF-cycles. But it may also happen that the clock readings in the fair zones in an NZF-cycle are upwardly bounded. In this case, we can resort to the strongly connected component algorithm to search for the NZF-cycles. Since the existence of a fair zone (one that satisfies $\eta_2$) is a necessary condition for NZF-cycles, we can start the search right from fair zones. The following procedure returns a fair zone in an NZF-cycle along which the cycle time is no less than 1.

```
get_a_zone_w_DFS(η₁, η₂) {
    While η₁ ∧ η₂ is not empty, do {
        Find a zone ζ in η₁ ∧ η₂;
        Use depth-first search in η₁ to find a path that ............................(3)
            • both starts and ends at ζ; and
            • the path time is no less than 1.
        If the path in statement (3) exists, return ζ; else η₁ := η₁ ∧ ¬ζ;
    }
    return false;
}
```

### 8.4 Algorithm for successive under-approximation

The formulation in lemma 5 does not suggest any particular methods to manage the evaluation process. However, if we have a policy to generate zone descriptions $\eta_1, \ldots, \eta_n$ one by one in succession, then we can evaluate $NZF(\eta_0, \eta_1, \eta_2)$ with successive under-approximation. In the following, we use lemma 5 and the two guessing techniques in subsections 8.2 and 8.3 to construct an algorithm to embody the idea. In particular, we use the number of enumeration iterations as the *level of approximation* to control the resource consumption and evaluation precision. The higher level of approximation is demanded, the more resources is consumed and the better evaluation precision could be achieved.

```
NZF_successive_uapprox(η₀, η₁, η₂, level)  /* level ≥ 0 */ {
    η := rch_bck(η₀, rch_bck(η₁, get_zones_wo_upperbounds(η₁ ∧ η₂))); η₁ := η₁ ∧ ¬η;
    for (l := 1; l ≤ level; l := l + 1) {
        ζ := get_a_zone_w_DFS(η₁, η₂); η := η ∨ rch_bck(η₀, rch_bck(η₁, ζ)); η₁ := η₁ ∧ ¬ζ;
    }
    return η;
}
```

We choose to use `get_zones_wo_upperbounds()` to generate the starting NZF-zones at level 0 because intuitively, it is in general much less expensive to run `get_zones_wo_upperbounds()` than `get_a_zone_w_DFS()`.

## 9 speed-up in iterative searches for NZF zones

After each iteration of procedure `NZF_successive_uapprox()` in subsection 8.4, we remove an NZF-zone from the search space to avoid redundant searching. This removal takes place at the end of lines 2 and 4 in the procedure. It is possible to prune the search space in bigger chunks. Specifically, two NZF-zones that belong to the same NZF-cycle may be used in two iterations to start the search in `NZF_successive_uapprox()`. Suppose the two zones are $\zeta, \zeta'$ used in this order. Then $\zeta' \subseteq \text{rch\_bck}(\eta_1, \zeta)$ and $\zeta \subseteq \text{rch\_bck}(\eta_1, \zeta')$. Thus there is no need to make a new round of depth-first search from $\zeta'$. In fact, we can establish the following lemma which can give us a sufficient condition leading to the significant pruning of the iterative search spaces.

**Lemma 6.** *Given three zones $\zeta_0$, $\zeta_1$, and $\zeta_2$ such that $\zeta_1$ and $\zeta_2$ are in the same $(\eta_1, \eta_2)$-NZF-cycle, if $\zeta_1 \subseteq \mathtt{rch\_bck}(\eta_1, \zeta_0)$, then $\zeta_2 \subseteq \mathtt{rch\_bck}(\eta_1, \zeta_0)$.* ∎

Proof of the lemma can be found in appendix B. Lemma 6 implies that we do not have to search through any states in $\mathtt{rch\_bck}(\eta_1, \zeta_0)$ after the iteration with $\zeta_0$ as the search start. This leads to the following revision of procedure $\mathtt{NZF\_successive\_uapprox}()$.

---

$\mathtt{NZF\_successive\_uapprox\_big\_chunks}(\eta_0, \eta_1, \eta_2, \mathit{level})$ /* $\mathit{level} \geq 0$ */ {
   $\eta := \mathtt{rch\_bck}(\eta_1, \mathtt{get\_zones\_wo\_upperbounds}(\eta_1 \wedge \eta_2));\ \eta_1 := \eta_1 - \eta;$
   $\eta := \mathtt{rch\_bck}(\eta_0, \eta);$
   for ($l := 1;\ l \leq \mathit{level};\ l := l + 1$) {
     $\zeta := \mathtt{rch\_bck}(\eta_1, \mathtt{get\_a\_zone\_w\_DFS}(\eta_1, \eta_2));\ \eta_1 := \eta_1 - \zeta;$
     $\eta := \eta \vee \mathtt{rch\_bck}(\eta_0, \zeta);$
   }
   return $\eta$;
}

---

## 10 Implementation and experiment

We have implemented the ideas in our model-checker/simulator, **RED** version 5.4, for communicating timed automatas [13]. **RED** uses the new BDD-like data-structure, *CRD* (Clock-Restriction Diagram) [16–18], and supports both forward and backward analyses, full *TCTL* model-checking with event constraints and multiple fairness assumptions [19], deadlock detection, and counter-example generation. We here report our experiments with the following seven parameterized benchmarks. The details of the benchmarks and their running options can be found in appendix C. The experiment result is shown in table 1. For each of the benchmarks, we have collected data with exact analysis and under-approximation in greatest fixpoint evaluation with or without the speed-up technique. For each run, we report the CPU time (for model-checking only), memory size, and the level of under-approximation needed for the verification. For all the benchmarks, our under-approximation techniques show significant enhancement over the exact analysis. For example, for benchmark (F), the performance can be a thousand times better.

Also the experiment outcome shows a good promise that our under-approximation formulation may be able to quickly refute specifications and prove vacuity. Most of the benchmarks can be verified with under-approximation of level zero or one.

| benchmarks | concurrency | exact (time/space) | under-approximation (time/space/level) | |
|---|---|---|---|---|
| | | | no speed-up | speed-up |
| (A) Fischer's bounded waiting | 2 proc's | 0.17s/30k | 0.03s/15k/0 | 0.02s/15k/0 |
| | 3 proc's | 2.68s/146k | 0.34s/73k/0 | 0.32s/73k/0 |
| | 4 proc's | 28.5s/580k | 2.27s/324k/0 | 2.24s/324k/0 |
| | 5 proc's | 275.1s/2656k | 14.71s/1335k/0 | 14.84s/1335k/0 |
| (B) Fischer's progress w. a bug | 2 proc's | 0.12s/21k | 0.04s/16k/1 | 0.04s/16k/1 |
| | 3 proc's | 1.93s/85k | 0.36s/60k/1 | 0.34s/60k/1 |
| | 4 proc's | 21.98s/354k | 2.75s/225k/1 | 2.83s/225k/1 |
| | 5 proc's | 215.0s/1543k | 22.46s/873k/1 | 22.27s/873k/1 |
| (C) CSMA/CD bounded waiting | 2 senders | 2.77s/116k | 0.48s/74k/2 | 0.44s/73k/1 |
| | 3 senders | 78.97s/942k | 4.58s/150k/5 | 3.12s/129k/1 |
| | 4 senders | 1560s/4536k | 31.10s/366k/2 | 27.89s/292k/1 |
| | 5 senders | > 8600s | 473.1s/1013k/2 | 214.9s/838k/1 |
| (D) CSMA/CD progress of retrials w. a bug | 2 senders | 0.32s/34k | 0.12s/34k/0 | 0.11s/34k/0 |
| | 3 senders | 4.47s/98k | 1.61s/69k/0 | 1.47s/69k/0 |
| | 4 senders | 33.28s/316k | 11.96s/186k/0 | 11.99s/186k/0 |
| | 5 senders | 220.9s/884k | 80.47s/596k/0 | 77.77s/596k/0 |
| (E) FDDI token ring vacuity | 2 stations | 2.88s/305k | 0.04s/42k/1 | 0.06s/42k/1 |
| | 3 stations | 152.15s/3108k | 0.29s/119k/1 | 0.30s/119k/1 |
| | 4 stations | > 1200s | 2.15s/617k/1 | 2.27s/618k/1 |
| | 5 stations | > 1200s | 70.78s/9574k/1 | 71.10s/9571k/1 |
| (F) PATHOS scheduling inevitable fairness | 2 proc's | 0.01s/10k | ≈ 0s/10k/0 | ≈ 0s/10k/0 |
| | 3 proc's | 0.22s/58k | 0.03s/24k/0 | 0.04s/24k/0 |
| | 4 proc's | 10.72s/1181k | 0.13s/61k/0 | 0.15s/61k/0 |
| | 5 proc's | 1280s/39076k | 0.60s/192k/0 | 0.61s/192k/0 |
| | 6 proc's | N/A | 2.17s/601k/0 | 2.10s/601k/0 |
| | 7 proc's | N/A | 6.44s/1813k/0 | 6.51s/1813k/0 |
| | 8 proc's | N/A | 20.19s/5253k/0 | 20.17s/5253k/0 |
| | 9 proc's | N/A | 62.02s/14632k/0 | 66.30s/14632k/0 |
| (G) Bluetooth | 9 procs | 228.9s/1886k | 35.98s/1280k/0 | 36.36s/1280k/0 |
| (H) L2CAP | 9 procs | 233.4s/1887k | 40.76s/1280k/0 | 41.21s/1280k/0 |
| (I) | 9 procs | 262.0s/1913k | 29.77s/1435k/0 | 29.71s/1435k/0 |
| (J) | 9 procs | 467.8s/2235k | 57.91s/1317k/0 | 56.93s/1317k/0 |
| (K) | 9 procs | 480.7s/2135k | 67.66s/1332k/0 | 71.56s/1332k/0 |
| (L) | 9 procs | 476.2s/2136k | 51.59s/1437k/0 | 52.97s/1437k/0 |
| (M) | 9 procs | 250.0s/1887k | 52.34s/1280k/0 | 52.89s/1280k/0 |
| (N) | 9 procs | 251.9s/1888k | 60.12s/1281k/0 | 55.19s/1281k/0 |
| (O) | 9 procs | 280.6s/1913k | 44.59s/1434k/0 | 44.98s/1434k/0 |

data collected on a Pentium 4 Mobile 1.6GHz with 256MB memory running LINUX;
s: seconds; k: kilobytes of memory in data-structure; N/A: not available;

**Table 1.** Performance data of model-checking algorithms

## 11   Conclusion

We investigate how to use under-approximation to fast refute incorrect inevitabilities and to check vacuous satisfaction in dense-time models. Experiment results showed that our techniques had significantly enhanced the performance of our model-checker against several benchmarks. In the future, we feel that such techniques could be useful in industrial projects.

# APPENDIX

## A  Proof for lemma 5

We first assume that there is a state $\nu \models NZF(\eta_0, \eta_1, \eta_2)$. This means that there are two states $\nu_1$ and $\nu_2$ in a run such that (1) $\nu_1$ and $\nu_2$ belong to the same zone; (2) $\nu_1$ and $\nu_2$ both satisfy $\eta_2$; (3) zones along the run segment from $\nu_1$ to $\nu_2$ all satisfy $\eta_1$; (4) time-passage of the run segment is no less than 1; and (5) the run segment from $\nu$ to $\nu_1$ is concatenated from two run segments such that the states in the first and the second segments respectively satisfy $\eta_0$ and $\eta_1$. Then the finite run segment from $\nu_1$ to $\nu_2$ can be projected to an $(\eta_1, \eta_2)$-NZF-cycle. Thus the zone of $\nu_1$, say $\zeta'$, is an $(\eta_1, \eta_2)$-NZF-zone. Then condition (5) in the above implies that there is an $\eta_0$-path concatenated with an $\eta_1$-path from the zone of $\nu$ to $\zeta'$. Thus we know that $\nu \models \texttt{rch\_bck}(\eta_0, \texttt{rch\_bck}(\eta_1, \zeta'))$.

Now we assume that there is a state $\nu \models \texttt{rch\_bck}(\eta_0, \texttt{rch\_bck}(\eta_1, \zeta'))$ for a particular $(\eta_1, \eta_2)$-NZF-zone $\zeta'$. This means that there is an $\eta_0$-path from the zone of $\nu$ to another $\zeta_1$ which in turn starts an $\eta_1$-path to the $(\eta_1, \eta_2)$-NZF-cycle which $\zeta'$ belongs to. These $\eta_0$-path, $\eta_1$-path, and the $(\eta_1, \eta_2)$-NZF-cycle together can be unrolled to an infinite zone path along which $\eta_0$ is true until $\eta_1$ becomes true forever and $\eta_2$ is true infinitely often. Since this unrolled infinite zone path is constructed in a backward analysis and $\nu$ is in the zone starting the path, we deduce that $\nu$ also starts a run that satisfies $\eta_0$ until $\eta_1$ is true forever and $\eta_2$ is true infinitely often. This means $\nu \models NZF(\eta_0, \eta_1, \eta_2)$.

## B  Proof for lemma 6

We need to prove that there is an $\eta_1$-path $P$ from $\zeta_2$ to $\zeta_0$. Since $\zeta_1 \subseteq \texttt{rch\_bck}(\eta_1, \zeta_0)$, there is an $\eta_1$-path $P_1$ from $\zeta_1$ to $\zeta_0$. Since $\zeta_1$ and $\zeta_2$ are in the same $(\eta_1, \eta_2)$-NZF-cycle, there is also an $\eta_1$-path $P_2$ from $\zeta_2$ to $\zeta_1$. Thus an example of $P$ is the concatenation of $P_2$ and $P_1$.

## C  Experiment and benchmarks

Our tool **RED** can be downloaded for free at http://cc.ee.ntu.edu.tw/∼val. To run the tool, simply type "`red [-options] ⟨inputfilename⟩ ⟨outputfilename⟩`" in LINUX. To invoke the under-approximation, please use option '`-Au`'; and over-approximation, please use option '`-Ao`.'

To enforce the non-Zeno requirement on computations, please use use option '`-Z`.'

To invoke greatest under-approximation of level d, please use option '`Gd`.' Note that if this option is invoked while the subformula is to be evaluated with over-approximation, then exact analysis will instead be carried out.

To invoke the speed-up technique, please use option '`Gc`.'

While we describe the following benchmarks, we shall also list the options used.

(A) *Bounded waiting of Fischer's mutual exclusion algorithm.* The algorithm uses a global pointer variable and a local clock of each process to guarantee the mutual exclusion to the critical section. The algorithm does not guarantee that a process in the `ready` state will eventually enter the critical section. We want to refute the following specification $\forall\Box(\texttt{ready}_1 \to \forall\Diamond\texttt{critical}_1)$ that is not true of the algorithm. Here $\texttt{ready}_i$ and $\texttt{critical}_i$ are the propositions respectively marking that process $i$ is in the `ready` and the `critical` states.

The input file names are like `hf`$i$`tb.d` where $i$ is the number processes. For exact analysis, we use option '`-Z`.' For under-approximation, we use options '`-AoGd0` (without the speed-up technique) and '`-AoGcGd0`' (with the speed-up technique).

(B) *Progress of Fischer's algorithm with a bug.* We inserted a bug in the Fischer's mutual exclusion algorithm so that processes may prevent each other from the critical section. We want to refute the following property
$$\forall\Box(\texttt{ready}_1 \to \forall\Diamond\exists i, \texttt{critical}_i)$$
saying that if a process wants to enter the critical section, then eventually some process will be in the critical section. The bug invalidates the property. The input file names are like `hf`$i$`err.d` where $i$ is the number processes. For exact analysis, we use option '`-Z`.' For under-approximation, we use options '`-AoGd1` (without the speed-up technique) and '`-AoGcGd1`' (with the speed-up technique).

(C) *Bounded waiting of CSMA/CD mutual exclusion algorithm.* CSMA/CD assumes that all senders share the same bus. Each of them can send a message to the bus when it finds no signals in the bus. But if in $52\mu s$ it finds its message has been corrupted, then it has to stop sending the message and retry later. We want to refute the following bounded waiting property
$$\forall\Box(\texttt{transm}_1 \to \forall\Diamond(\texttt{transm}_1 \land x_1 \geq 52))$$
that is not guaranteed in CSMA/CD. Here proposition $\texttt{transm}_i$ means that process $i$ is sending out a message and $x_i$ is the local clock of process $i$.

The input file names are like `hcd`$i$`af.d` where $i$ is the number processes. For exact analysis, we use option '`-Z`.' For under-approximation without speed-up, we use options '`-AoGd2`' for all files except `hcd3af.d` and '`AoGd5`' for `hcd3af.d`. With speed-up, we use '`-AoGcGd1`.'

(D) *Progress of retrials of CSMA/CD mutual exclusion algorithm with a bug.* In the CSMA/CD model, processes will try to resend the messages until they succeed in sending out the messages. We inserted a bug to the algorithm so that some processes may be trapped in an error mode with no outgoing transitions. We want to refute the property $\forall\Box(\texttt{retry}_1 \to \forall\Diamond\exists i, \texttt{transm}_i)$ that is not guaranteed in CSMA/CD with this bug. Here $\texttt{retry}_i$ means that process $i$ is waiting for resending its message.

The input file names are like `hcd`$i$`err2.d` where $i$ is the number processes. For exact analysis, we use option '`-Z`.' For under-approximation, we use options '-AoGd0 (without the speed-up technique) and '-AoGcGd0' (with the speed-up technique).

(E) *Vacuity of a strong fairness assumption in FDDI token ring protocol.* We want to check whether there is a run along which a station enters the asynchronous transmission mode infinitely many times. That is $\exists\Box\Diamond\mathtt{async}_1$.

The input file names are like hddi*i*aeo.d where *i* is the number processes. For exact analysis, we use option '-Z.' For under-approximation, we use options '-AuGd1 (without the speed-up technique) and '-AuGcGd1' (with the speed-up technique).

(F) *Strong fairness for the lowest priority in PATHOS real-time operating system scheduling policy.* PATHOS is a real-time operating system that uses priority scheduling policy. In a system of $n$ processes, the model assumes that in each period of $n$ time units, each process will need at most 1 time units to run on the CPU. We want to refute the specification that the lowest-priority process gets to run on the CPU infinitely often along any computation. That is, $\forall\Box\Diamond run_n$.

The input file names are like pathos*i*aao.d where *i* is the number processes. For exact analysis, we use option '-Z.' For under-approximation, we use options '-AoGd0 (without the speed-up technique) and '-AoGcGd0' (with the speed-up technique).

(G-O) *Various properties of Bluetooth L2CAP.* We have also checked 9 properties against Bluetooth L2CAP []. The L2CAP model is refined from the one in []. There are 9 processes to model the users, L2CAP layers, timers, and the medium in the two sides of the communications. We checked the model against the following the following 9 properties. For convenience, we use '($\alpha$),' '($\beta$),' and '($\gamma$)' to represent the following properties respectively: ($\alpha$): the master L2CAP is in state 'W4_L2CAP_CONNECT_RSP; ($\beta$): the master L2CAP is in state 'W4_L2CA_CONNECT_RSP; and ($\gamma$): the master L2CAP is in state 'OPEN.' The nine properties and their input file names are labeled as follows:

| Labels | properties | input file name |
| --- | --- | --- |
| (G) | $\forall\Box((\alpha) \to \forall\Diamond(\gamma))$ | l2nae.d |
| (H) | $\forall\Box((\alpha) \to \forall\Diamond^\infty(\gamma))$ | l2nao.d |
| (I) | $\forall\Box((\alpha) \to \forall\Box(\gamma))$ | l2nag.d |
| (J) | $\forall\Box((\alpha) \to \forall\Diamond((\beta) \land \forall\Diamond(\gamma)))$ | l2naeae.d |
| (K) | $\forall\Box((\alpha) \to \forall\Diamond((\beta) \land \forall\Diamond^\infty(\gamma)))$ | l2naeao.d |
| (L) | $\forall\Box((\alpha) \to \forall\Diamond((\beta) \land \forall\Box(\gamma)))$ | l2naeag.d |
| (M) | $\forall\Box((\alpha) \to \forall\Box((\beta) \to \forall\Diamond(\gamma)))$ | l2nagae.d |
| (N) | $\forall\Box((\alpha) \to \forall\Box((\beta) \to \forall\Diamond^\infty(\gamma)))$ | l2nagao.d |
| (O) | $\forall\Box((\alpha) \to \forall\Box((\beta) \to \forall\Box(\gamma)))$ | l2nagag.d |

For exact analysis, we use option '-Z.' For under-approximation, we use options '-AoGd0 (without the speed-up technique) and '-AoGcGd0' (with the speed-up technique).